\documentstyle[12pt,epsf]{article}
\def\beq{\begin{equation}}
\def\eeq{\end{equation}}
\def\bea{\begin{eqnarray}}
\def\eea{\end{eqnarray}}
\def\bem{\begin{math}}
\def\eem{\end{math}}
\def\bit{\begin{itemize}}
\def\eit{\end{itemize}}
\def\bla{\begin{flushright}}
\def\ela{\end{flushright}}
\def\qq2{$Q^2$}               
\def\aa1{$A_1(x,Q^2)$}        
\def\ff1{$F_1(x,Q^2)$}        
\def\gg1{$g_1(x,Q^2)$}        
\begin{document}
\vskip 3cm
\begin{center}
{\Large{\bf About the $Q^2$ dependence of asymmetry $A_1$}}
\end{center}
\vskip 1.7cm
\begin{center}
{\bf A.V.Kotikov}
\footnote{~On leave of absence from Particle Physics Laboratory, JINR, 
Dubna, Russia.\\
~E-mail: kotikov@lapphp0.in2p3.fr; kotikov@sunse.jinr.dubna.su}
\\
{\it Laboratoire de Physique Theorique ENSLAPP\\
LAPP, B.P. 110, F-74941, Annecy-le-Vieux Cedex, France}\\
and\\ {\bf D.V.Peshekhonov}
\footnote{~E-mail:
dimitri@na47sun05.cern.ch;
peshehon@sunse.jinr.dubna.su}
\\
{\it Particle Physics Laboratory, JINR, Dubna, Russia.}
%
\end{center}
\vskip 3cm
{\large{\bf Abstract}}\\
\vskip .5cm \hskip -.56cm
We analyse the proton and deutron data on spin dependent asymmetry
~\aa1 supposing the DIS structure functions $g_1(x,Q^2)$ 
and $F_3(x,Q^2)$ have the similar $Q^2$-dependence. 
As a result, we have obtained $\Gamma_1^p - \Gamma_1^n = 0.192$ at 
$Q^2= 10~{\rm GeV}^2$ and $\Gamma_1^p - \Gamma_1^n = 0.165$ at 
$Q^2= 3~{\rm GeV}^2$, in the best agreement with the Bjorken 
sum rule predictions.\\
PACS number(s): 13.60.Hb, 11.55.Hx, 13.88.+e \\
\newpage
%
%
\hskip -.56cm
An experimental study of the nucleon spin structure is realized by 
measuring of the asymmetry $A_1(x,Q^2) = g_1(x,Q^2) / F_1(x,Q^2)$. 
The most known theoretical predictions on spin dependent structure 
function $g_1(x,Q^2)$ of the nucleon were done by Bjorken \cite{Bj} and 
Ellis and Jaffe \cite{EJ} for the so called {\it first moment value} 
$\Gamma_1 = \int_0^1 g_1(x) dx$.\\
The calculation of the $\Gamma_1$ value requires the knowledge of 
structure function $g_1$ at the same $Q^2$ in the hole $x$ range. 
Experimentally asymmetry $A_1$ is measuring at different values of $Q^2$ 
for different $x$ bins. 
An accuracy of the past and modern experiments 
\cite{E80} - \cite{E143d} 
allows to analyze data in the assumption \cite{EK93}-\cite{Close}
that asymmetry \aa1 is \qq2 
independent (structure functions $g_1$ and $F_1$ have the same $Q^2$ 
dependence). But the tune checking of the Bjorken and Ellis - Jaffe sum 
rules requires considering the $Q^2$ dependence of $A_1$ or $g_1$ 
(for recent studies of the $Q^2$ dependence of $A_1$  see 
\cite{EK93}-\cite{E143pd})
%
\vskip .4cm \hskip -.56 cm
This article is based on our observation\footnote{The conclusion
  connects with
  our previous analysis \cite{KoPe}.} 
that the $Q^2$ dependence of $g_1$ 
and the spin average structure function $F_3$ is the same in a wide $x$ range: 
$10^{-2} < x < 1$. At small $x$ it seems that may be not true (see 
\cite{Gluck}, \cite{BFR}-\cite{EMR}).\\

\hskip -.56cm
To demonstrate the validity of the observation, 
lets consider the nonsinglet (NS) $Q^2$ evolution of structure functions 
$F_1,~g_1$ and $ F_3$. The DGLAP equation for the NS part of these functions can be presented as\footnote{We use $\alpha(Q^2)= \alpha_s(Q^2)/{4 \pi}$ .} :\\
\bea
{dg_1^{NS}(x,Q^2) \over dlnQ^2} = -{1 \over 2} \gamma_{NS}^-(x, \alpha) 
\times g_1^{NS}(x,Q^2),
\nonumber \\
{dF_1^{NS}(x,Q^2) \over dlnQ^2} = -{1 \over 2} \gamma_{NS}^+(x, \alpha) 
\times F_1^{NS}(x,Q^2),
\label{1} \\
{dF_3^{NS}(x,Q^2) \over dlnQ^2} = -{1 \over 2} \gamma_{NS}^-(x, \alpha) 
\times F_3^{NS}(x,Q^2),
\nonumber 
\eea
where symbol $\times$ means the Mellin convolution. 
Functions $\gamma^{\pm}_{NS}$ are the reverse Mellin transforms of the 
anomalous dimensions $\gamma^{\pm}_{NS}(n, \alpha)= \alpha
\gamma^{(0)}(n)_{NS}  + 
\alpha^2 \gamma^{\pm (1)}_{NS}(n) + O(\alpha^3)$ and the Wilson
coefficients\footnote{Because we consider here the 
structure functions themselves 
but not the quark distributions. Note that
more standard definition of $b^{+}_{NS}(n)$ and
$b^{-}_{NS}(n)$ are $b_{1,NS}(n) = b_{2,NS}(n) - b_{L,NS}(n)$ and
$b_{3,NS}(n)$.}  
$\alpha b^{\pm}(n) + O(\alpha^2)$ :
 \begin{eqnarray} 
\gamma^{ \pm}_{NS}(x,\alpha) ~=~ \alpha 
  \gamma^{(0)}_{NS}(x) + \alpha^2 \biggl(
  \gamma^{\pm (1)}_{NS}(x) +
  2\beta_0 b^{\pm}(x)  \biggr) + O(\alpha ^3),  
\label{2}  \end{eqnarray}
where $\beta (\alpha)= - \alpha^2 \beta_0 - 
 \alpha^3 \beta_1 + O(\alpha^4)$ is QCD $\beta$-function.\\
The above mentioned Mellin transforms mean that
 \begin{eqnarray} 
f(n,Q^2) ~=~ \int^1_0 dx x^{n-1} f(x,Q^2),
\label{3}  \end{eqnarray}
where $ f= \{\gamma^{(0)}_{NS}, \gamma^{\pm (1)}_{NS}, b^{\pm}_{NS},
\gamma^{(k)}_{ij}, \gamma^{* (k)}_{ij}, b_{i} \mbox{ and } b^{*}_{i} \} $
with $k = 1,2$ and $\{i,j \} = \{S,G \}$. \\
Eqs. (1) show the $Q^2$ dependence of NS parts of $g_1$ and $F_3$ is the same 
(at least in first two orders of the perturbative QCD \cite{Kodaira})
and differs from $F_1$ already in the first subleading order 
($\gamma^{+(1)}_{NS} \neq \gamma^{-(1)}_{NS}$ \cite{RoSa} and
$b^+_{NS} - b^-_{NS} = (8/3)x(1-x)$).\\
For the singlet parts of $g_1$ and $F_1$ evolution equations are :\\
\bea 
{dg_1^S(x,Q^2) \over dlnQ^2} &= -{1 \over 2} \biggl[
\gamma_{SS}^{*}(x, \alpha) \times g_1^S(x,Q^2) +
\gamma_{SG}^{*}(x, \alpha) \times \Delta G(x,Q^2) \biggr],
\nonumber
\\
{dF_1^S(x,Q^2) \over dlnQ^2} &= -{1 \over 2} \biggl[
\gamma_{SS}(x, \alpha) \times F_1^S(x,Q^2) +
\gamma_{SG}(x, \alpha) \times G(x,Q^2) \biggr],
\label{4} 
\eea
where
 \bea
\gamma_{SS}(x,\alpha) = \alpha 
  \gamma^{(0)}_{SS}(x) + \alpha^2 \biggl(
  \gamma^{(1)}_{SS}(x) +  b_G(x) \times \gamma^{(0)}_{GS}(x) +
  2\beta_0 b_S(x)  \biggr) 
\nonumber \\
+ O(\alpha ^3),  
\nonumber  \\
\gamma_{SG}(x,\alpha) = \frac{e}{f} \biggl[ \alpha 
  \gamma^{(0)}_{SG}(x) + \alpha^2 \biggl(
  \gamma^{(1)}_{SG}(x) + b_G(x) \times \bigl( 
  \gamma^{(0)}_{GG}(x) - \gamma^{(0)}_{SS}(x) \bigr) 
+ 2\beta_0 b_G(x) 
\nonumber \\
+  b_S(x) \times \gamma^{(0)}_{SG}(x) 
\biggr)
  \biggl] + O(\alpha ^3) \nonumber  \eea 
where $e = \sum_i^f e^2_i$ is
the sum of charge squares of $f$ active quarks.
The equations for polarized anomalous dimensions
$\gamma_{SS}^{*}(x, \alpha)$ and $\gamma_{SG}^{*}(x, \alpha)$
are similar. They ban be obtained by replacing 
$\gamma^{(1)}_{Si}(x) \to \gamma^{*(1)}_{Si}(x)$ and $b_i(x) \to 
b^*_i(x)$ ($i=\{S,G \}$).\\

Note here the gluon term is not negligible for $F_1$ at $x < 0.1$ but for  
$g_1$ we can neglect the gluons for $x>0.03$
\cite{GS95}-\cite{BFR}. 
The value $b^*_s(x)$ ($b_s(x)$) coincides with $b^-(x)$ ($b^+(x)$).
The difference between 
$\gamma_{NS}^{-(1)}$ and $\gamma_{SS}^{* (1)}+  b^*_G(x) \times
\gamma^{(0)}_{GS}(x)$  is negligible due to 
its difference 
having no a power singularity at $x \to 0$ 
(i.e. no a singularity for them momentum transforms 
at $n \to 1$ in momentum space)
and decreases as $O(1-x)$ at $x \to 1$ 
\cite{MeNe} (see also \cite{WVo}).
Contrary to this, the difference between
$\gamma_{SS}^{(1)}+  b_G(x) \times \gamma^{(0)}_{GS}(x)$ and
$\gamma_{SS}^{* (1)}+  b^*_G(x) \times \gamma^{(0)}_{GS}(x)$
contains the power singularity at $x \to 0$ (see \cite{GLY, Kodaira}). 
\\
This observation allows us to conclude the function :
\bea
A_1^*(x) = {g_1(x,Q^2) \over F_3(x,Q^2)}
\label{4.1}
\eea
should be practically $Q^2$ independent at $x>0.01$.
%
Because the r.h.s. of Eqs.(\ref{1}) and (\ref{4}) contain integrals of
structure functions, the approximate validity of (\ref{4.1})
is supported also by the same $x$-dependence of $g_1(x,Q^2)$
and $F_3(x,Q^2)$ at fixed $Q^2$. \\
The asymmetry $A_1$ at $Q^2=<Q^2>$ can be defined than as :
\bea
A_1(x_i,<Q^2>) =  {F_3(x_i,<Q^2>) \over F_3(x_i,Q^2_i)} \cdot
{F_1(x_i,Q^2_i) \over F_1(x_i,<Q^2>)} \cdot A_1(x_i,Q^2_i),
\label{5}
\eea
where $x_i$ ($Q^2_i$) means an experimentally measured value of $x$ ($Q^2$).\\
%
%
\vskip .4cm \hskip -.56cm
We use SMC and E143 proton and deuteron data for asymmetry $A_1(x,Q^2)$ 
\cite{SMCp} - \cite{E143d}. To get $F_1(x,Q^2)$ we take NMC parametrization 
for $F_2(x,Q^2)$ 
\cite{NMC} 
and SLAC parametrization for $R(x,Q^2)$ \cite{SLAC} ($F_1 \equiv F_2/2x[1+R]$). 
To get the values of $F_3(x,Q^2)$ we parametrize the CCFR data \cite{CCFR} as 
a function of $x$ and $Q^2$ (see Fig.1). \\
First, using eq.5, we recalculate the SMC \cite{SMCp,SMCd} and E143 
\cite{E143p,E143d} measured asymmetry of the 
proton and deuteron at $Q^2= 10~ {\rm GeV}^2$ and $3~ {\rm GeV}^2$, which are 
average $Q^2$ of these experiments respectively (results are shown in Fig.2, 3) 
and get the value of $\int g_1(x) dx$ through the measured $x$ ranges 
(see Table 1).\\
To obtain the first moment values $\Gamma_1^{p(d)}$ we have used an original 
estimations of SMC and E143 for unmeasured regions \cite{SMCp} - \cite{E143d}. Results on the 
$\Gamma_1$ values are shown in the Table 1.\\

\hskip -.56cm
{\bf Table 1.}~~ The first moment value of $g_1$ of the proton and deuteron.\\
\vskip -.9cm 
\begin{table}[h]
\begin{center}
\begin{tabular}{|c|c|c|c|c|c|}
\hline \hline
                     &         &             &          &                 &  \\
$x_{min} -- x_{max}$ & $<Q^2>$ & target      & $\int_{x_{min}}^{x_{max}} g_1 dx$ & 
$\Gamma_1$ & experiment \\
                     &         & type        &          &                 &  \\
\hline \hline
.003 -- 0.7 & $10~{\rm GeV}^2$ & proton      & 0.130 & 0.135 &SMC \\
.003 -- 0.7 & $10~{\rm GeV}^2$ & deuteron    & 0.038 & 0.0362 &SMC \\
\hline
.029 -- 0.8 & $3~{\rm GeV}^2$ & proton      & 0.123 & 0.130 &E143 \\
.029 -- 0.8 & $3~{\rm GeV}^2$ & deuteron      & 0.043 & 0.044 &E143 \\
\hline
\hline
\end{tabular}
\end{center}
\end{table}
\vskip -.4cm \hskip -.56cm
As the last step we calculate the difference 
$\Gamma_1^p - \Gamma_1^n = 2 \Gamma_1^p - ( \Gamma_1^p + \Gamma_1^n)$ where 
$\Gamma_1^p + \Gamma_1^n = 2 \Gamma_1^d / (1-1.5 \cdot \omega_D)$ and 
$\omega_D=0.05$ \cite{SMCd,E143d} .\\ 
At $Q^2=10~ {\rm GeV}^2$ we get the following results :\\
\bea
\Gamma_1^p - \Gamma_1^n = 0.199 \pm 0.038~~~~~~~~~~~~~ \mbox{(SMC \cite{SMCd})}
\nonumber \\
\Gamma_1^p - \Gamma_1^n = 0.192~~~~~~~~~~~~~~~~~~~~~ \mbox{(our result)}
\\
\Gamma_1^p - \Gamma_1^n = 0.187 \pm 0.003~~~~~~~~~~~~~~ \mbox{(Theory)}
\nonumber
\eea 
and at  $Q^2=3~ {\rm GeV}^2$ :
\bea
\Gamma_1^p - \Gamma_1^n = 0.163 \pm 0.026~~~~~~~~~~~~~ \mbox{(E143 \cite{E143d})}
\nonumber \\
\Gamma_1^p - \Gamma_1^n = 0.165~~~~~~~~~~~~~~~~~~~~~ \mbox{(our result)}
\\
\Gamma_1^p - \Gamma_1^n = 0.171 \pm 0.008~~~~~~~~~~~~~~ \mbox{(Theory)}
\nonumber
\eea 
\vskip .4cm \hskip -.56cm
As a conclusion, we would like to note 
\begin{itemize}
\item our observation that function $A_1^*(x)$ is $Q^2$ independent at 
large and intermediate $x$ 
is supported by good agreement (see Fig. 2,3) of
present analysis with other estimations \cite{ANR}-\cite{Gluck} of
the $Q^2$ dependence of the $A_1$;
\item 
at small $x$ structure functions $g_1(x,Q^2)$ and $F_3(x,Q^2)$ may have the 
same behaviour too (in traditional, Regge-motivated consideration 
$f  \sim x^{\delta}$, where $  \delta \geq 0)$ ) 
\cite{Heinmann} 
\footnote{According to the recent analysis \cite{BFR}-\cite{EMR}, however, the situation may be more complicated.}. 
\item The value of $\Gamma_1^p - \Gamma_1^n$ obtained in the supposion 
that $g_1$ and $F_3$ have the same $Q^2$ dependence improves the agreement 
with the Bjorken sum rule prediction.
\end{itemize} 

\hskip -.56cm
{\large \bf Acknowledgements}\\

\hskip -.56cm
We are grateful to W.G.~Seligman for 
providing us the available CCFR data of Ref.\cite{CCFR} and
A.V.~Efremov for discussions.\\
%
This work is supported partially by the Russian Fund for Fundamental Research, 
Grant N 95-02-04314-a.\\
%
%

%
\newpage
{\Large {\bf Figure Captions}}\\
\vskip 1cm
\begin{description} 
\item [Figure 1.]
DIS structure function $F_3(x,Q^2)$. CCFR data \cite{CCFR} and the 
parametrization.
\item [Figure 2.]
SMC \cite{SMCp} and E143 \cite{E143p} measured virtual photon-nucleon 
asymmetry $A_1^p(x)$~ as a function of $x$ (shown as a close points) 
in comparison with evolved to $Q^2$~= $10~(3)~{\rm GeV}^2$, respectively 
(open points).
\item [Figure 3.]
SMC \cite{SMCd} and E143 \cite{E143d} measured virtual photon-nucleon 
asymmetry $A_1^d(x)$~ as a function of $x$ (shown as a close points) 
in comparison with evolved to $Q^2$~= $10~(3)~{\rm GeV}^2$, respectively 
(open points).
\end{description}

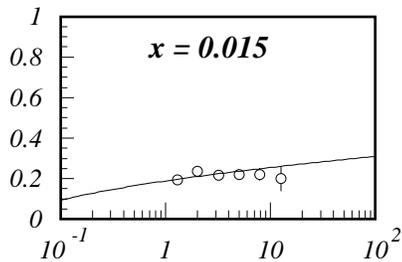
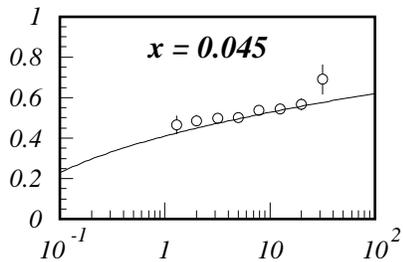
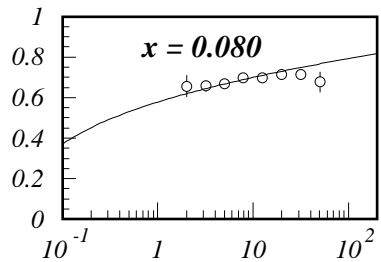
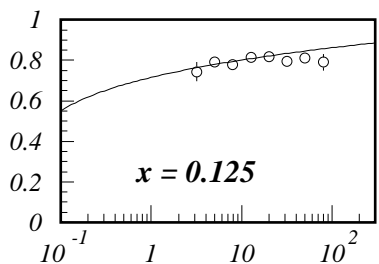
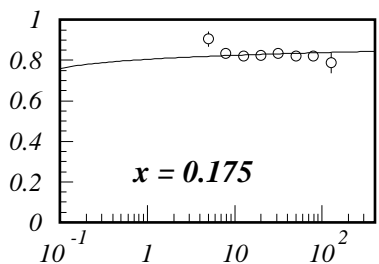
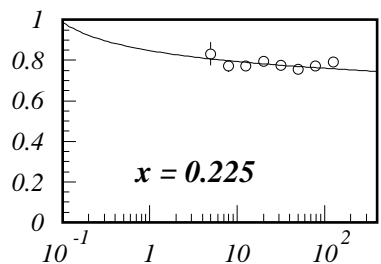
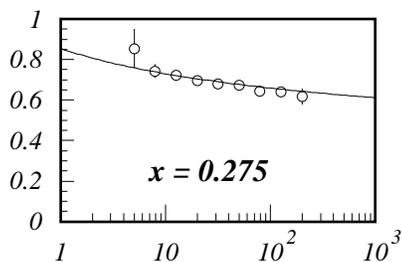
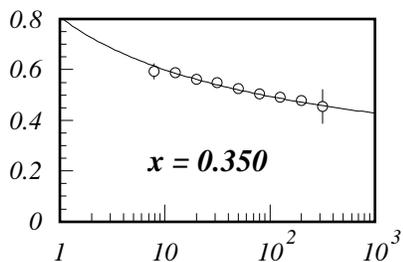
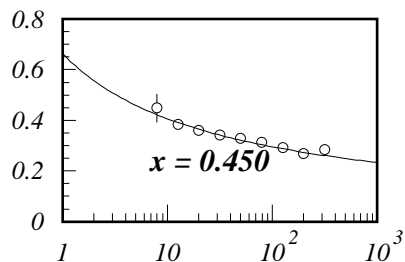
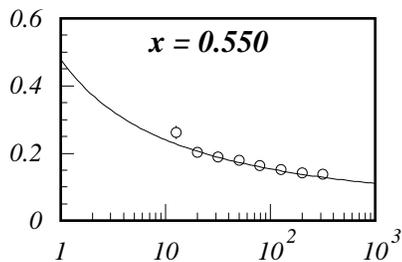
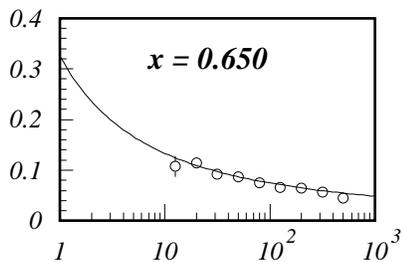
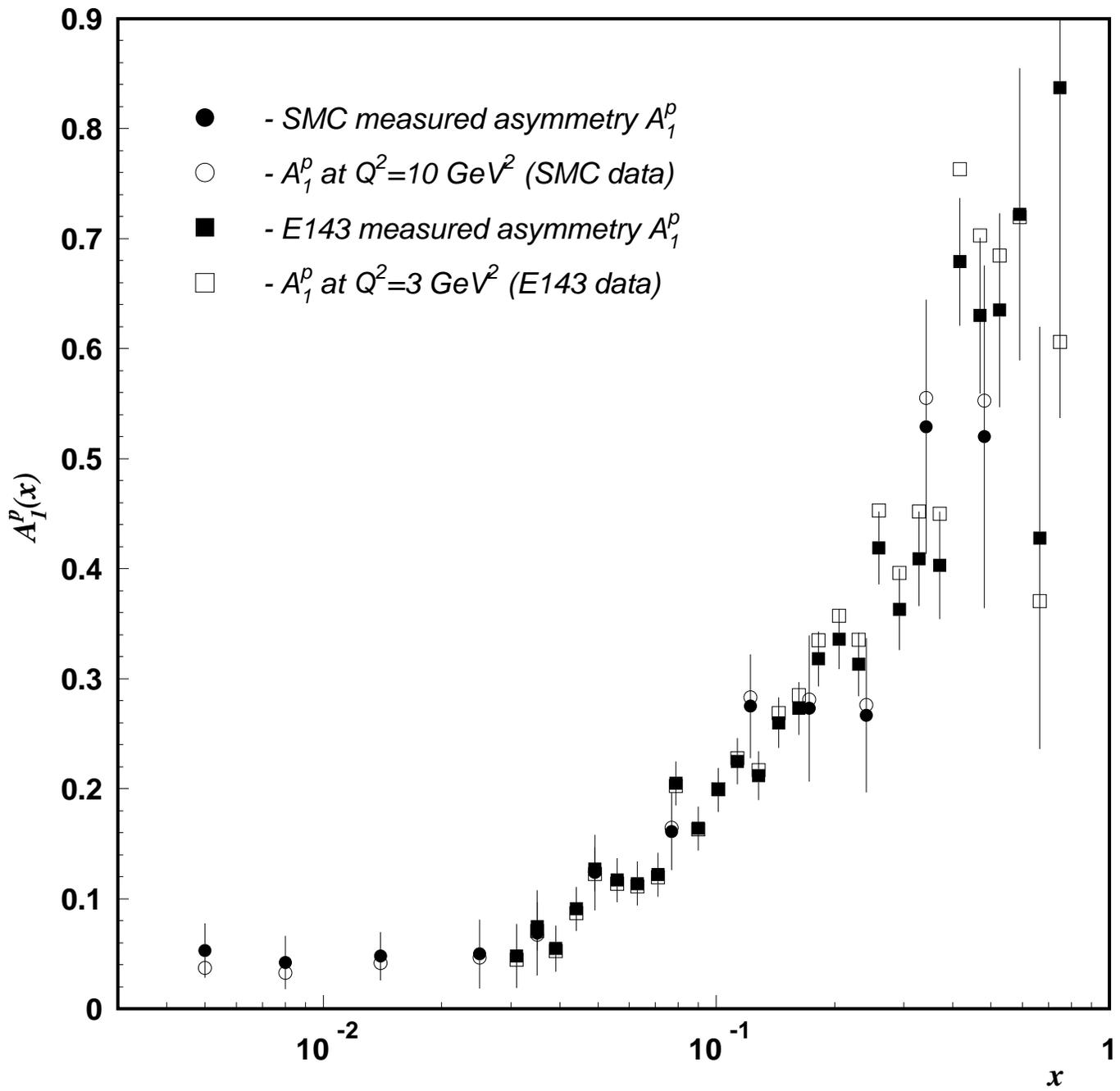
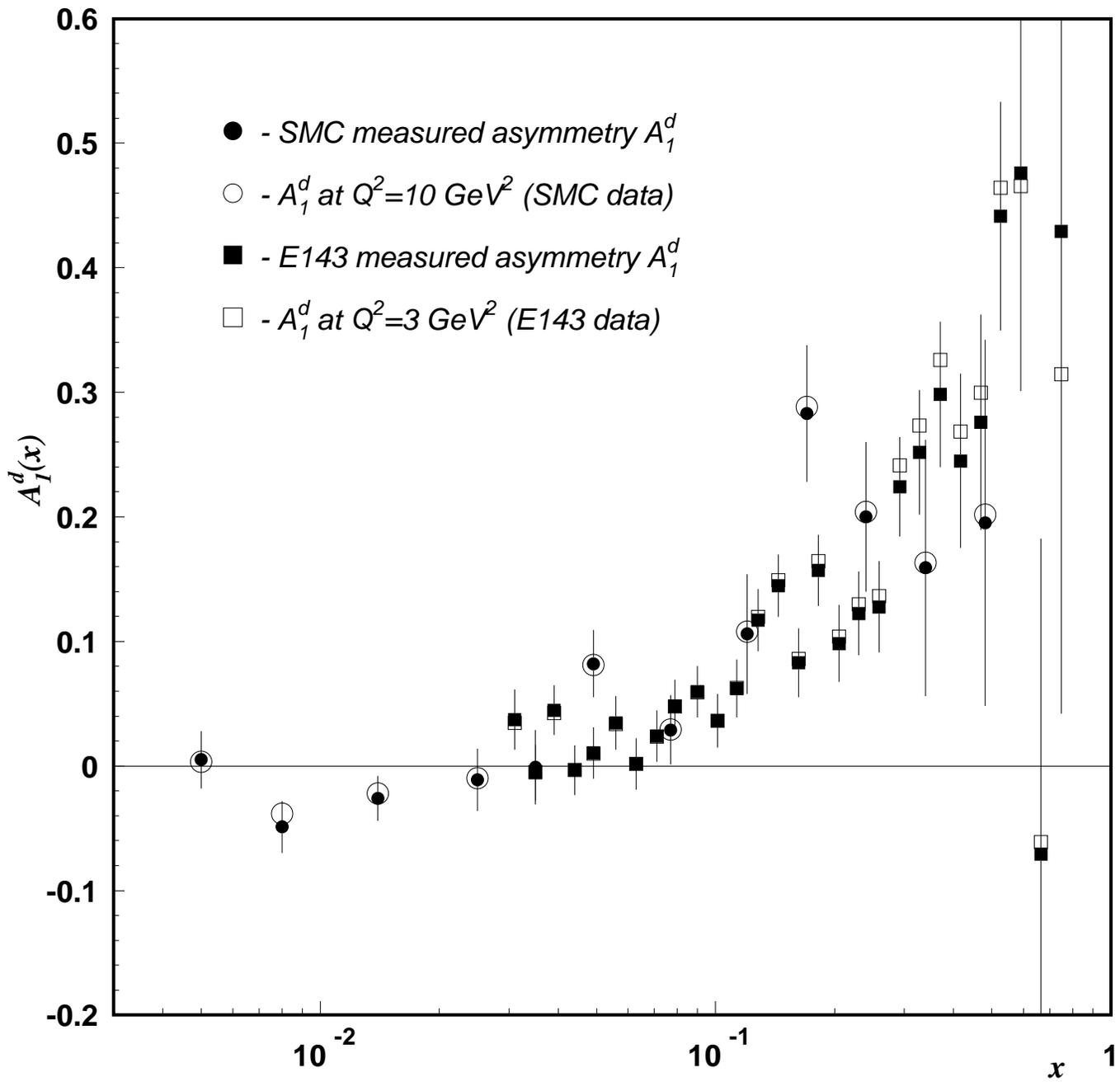
\end{document}